\documentclass[reprint, secnumarabic,amssymb, nobibnotes, aps, prl]{revtex4-1}

\usepackage{color}
\usepackage{epsfig, graphics, graphicx, subfigure, wrapfig, lipsum, longtable,array}
\usepackage{verbatim, import}
\usepackage{dcolumn}% Align table columns on decimal point
\usepackage{bm}% bold math
\usepackage{amsmath, braket}
\usepackage{array}
\usepackage{xr}
\newcolumntype{X}[1]{>{\centering\arraybackslash\hspace{0pt}}p{#1}}
\newcolumntype{M}[1]{ >{\centering\arraybackslash}m{#1}}
\newcommand{\roml}[1]{\lowercase\expandafter{\romannumeral #1\relax}}
\newcommand{\romu}[1]{\uppercase\expandafter{\romannumeral #1\relax}}

\begin{document}

\title{
Single-channel or multi-channel thermal transport? Effect of higher-order anharmonic corrections on the  predicted phonon thermal transport properties of semiconductors
}

\author{Ankit Jain}
\email{a\_jain@iitb.ac.in}
\affiliation{Mechanical Engineering Department, IIT Bombay, India}
\date{\today}%

\begin{abstract}
The phonon thermal transport properties of eight ternary intermetallic semiconductors are investigated by accounting for higher-order four-phonon scattering, phonon renormalization, and multi-channel thermal transport. The commonly used lowest-order theory, which accounts only for three-phonon scattering and without phonon renormalization, fails drastically for considered materials and underpredicts the thermal conductivity by up to a factor of two. The thermal conductivity decreases for three compounds and increases for five compounds with the application of higher-order corrections owing to a contrasting role of four-phonon scattering and phonon stiffening on the predicted thermal conductivity. %The weighted three-phonon scattering phase-space and effective bond stiffness are identified as simple material descriptors capable of characterizing the role of higher-order theory on predicted thermal conductivity without carrying out detailed calculations. 
Using the higher-order theory, at a temperature of 300 K, the lowest obtained thermal conductivity is $0.31$ W/m-K for $\text{BiCsK}_2$  and three other compounds ($\text{SbCsK}_2$, $\text{SbRbNa}_2$, and $\text{SbRbK}_2$) have thermal conductivities lower than $0.5$ $\text{W/m-K}$ via the particle-like phonon transport channel. The contribution from the wave-like coherent transport channel is lower than $0.05$ W/m-K in all of these ultra-low thermal conductivity  compounds. The higher-order theory is a must for the correct description of thermal transport physics, failing which the thermal transport is wrongly characterized as multi-channel transport by the lowest-order theory.
\end{abstract}
\maketitle

%\section{Introduction}
With advances in computational resources in the past decade, it is now possible to predict the phonon thermal conductivity of semiconducting crystalline materials by solving the Boltzmann transport equation (BTE) with inputs from ab-initio calculations \cite{esfarjani2008, esfarjani2011, lindsay2018, mcgaughey2019}. The BTE theory accounts only for three-phonon interactions at the lowest order \cite{lindsay2013a, lindsay2013b, mcgaughey2019}. While this lowest-order theory is found to work well for relatively simple materials and an excellent agreement with experiments is obtained wherever possible \cite{lindsay2013b, jain2015b,lindsay2018}, recently, this theory has been reported to fail severely for several low and high thermal conductivity compounds (such as BAs\cite{chen2020}, $\text{Tl}_3\text{VSe}_4$\cite{jain2020}, $\text{MoS}_2$\cite{gokhale2021}). In particular, it is shown that (a) phonon-scattering due to higher-order four-phonon processes and (b) phonon renormalization due to quartic anharmonicity are  significant in these technologically relevant materials not accounted for in the lowest-order theory \cite{ravichandran2018, feng2018, xia2020, ravichandran2020}. For binary compounds  with zinc blende crystal structure, applying the first of these  corrections resulted in up to 75\% reduction in thermal conductivity at room temperature \cite{ravichandran2020}. For $\text{Tl}_3\text{VSe}_4$, on the other hand, these higher-order corrections resulted in more than a factor of two increase in the thermal conductivity \cite{xia2020, jain2020}. In comparison to the lowest-order theory, the higher-order corrections are 3-4 orders of magnitude expensive to compute and their application is currently limited to handful of materials \cite{ravichandran2018, feng2018, xia2020, ravichandran2020, jain2020}.

In parallel, the failure of the lowest-order transport theory to describe the experimentally measured thermal transport in ultra-low thermal conductivity solids has also motivated the development of several heuristics-based multi-channel thermal transport models where on top of the particle-like phonon transport channel, an additional contribution arising from diffusion-like heat transfer channel is taken into account to explain the underprediction of experimentally measured values \cite{mukhopadhyay2018, luo2020}. These developments subsequently evolved into a derivation of unified thermal transport theory capable of describing the thermal transport in the spectrum of materials varying from purely amorphous to perfectly crystalline solids \cite{simoncelli2019}. The major finding of this unified thermal transport theory is the contribution from both particle-like (originating from well-defined phonon modes) and wave-like (from tunneling/coherence between modes) transport channels towards the thermal transport. In practical implementations, the  unified/multi-channel thermal transport theory requires phonon mode properties (frequency, linewidth, etc) which can be obtained by either using only the lowest-order theory or by also including the higher-order corrections. Since higher-order corrections are computationally demanding, the application of multi-channel thermal transport in literature is majorly limited to lowest-order theory driven phonon properties and its validity/significance is debated when higher-order corrections are taken into account \cite{xia2020}.

%As such, there is a need for a systematic understanding of these higher-order corrections for other material systems. More importantly, there is a need for development/identification of simple material descriptors which  are capable  of characterizing the  importance of these higher-order corrections on the predicted phonon thermal transport properties without requiring detailed calculations. Finally, there is also a need to understand if the thermal transport is via single transport channel or via multi-transport channel when correct phonon properties are used in the unified thermal transport theory. 

In this regard, the phonon thermal transport of eight ternary semiconducting compounds with stoichiometry $\text{ABC}_2$  [A $\in$ (Sb, Bi), B $\in$ (Na, K, Rb, Cs) and C $\in$ (Li, Na, K)] is investigated in this work by employing the higher-order corrections and unified thermal transport theory. At a temperature of 300 K, the thermal conductivity of these compounds varies between $0.16$ - $6.1$ $\text{W}/\text{m-K}$ using the lowest-order theory. The combined effect of four-phonon scattering and phonon renormalization resulted in a decrease in thermal conductivity for three of the considered compounds and an increase in thermal conductivity for five considered compounds. The maximum change in thermal conductivity is for $\text{SbRbK}_2$, for which thermal conductivity increases by more than a factor of two (from $0.16$ to $0.37$ $\text{W/m-K}$) with the inclusion of these corrections. %The weighted three-phonon scattering phase-space and effective bond stiffness are identified as simple and easy-to-compute material descriptors capable of characterizing the contrasting role of considered higher-order corrections on the thermal conductivity. 
The particle-like single-channel phonon transport picture is erroneously invalid at the lowest-order of theory and the application of higher-order corrections results in a single-channel thermal transport with coherent channel contribution less than 13\% in all of the considered compounds.

%\section{Methodology}
The contribution of particle-like phonon transport channel towards the thermal transport in the $\alpha$-direction  is obtained by solving  the Boltzmann transport equation and the Fourier's law, as \cite{reissland1973, dove1993}:
\begin{equation}
 \label{eqn_theory_conduct}
 k_{\alpha}^{p} = \sum_{{{\lambda}}}  c_{{{\lambda}}} v_{{{\lambda}}, \alpha}^{2} \tau_{{{\lambda}}, \alpha},
\end{equation}
where $k_{\alpha}^p$ represents the thermal conductivity, the summation is over all the phonon wavevectors, ${\boldsymbol{q}}$, and polarizations, $\nu$, enumerated by composite index $\lambda = ({\boldsymbol{q}}, \nu)$,
$c_{{{\lambda}}}$ is the phonon specific heat, $v_{{{\lambda}},\alpha}$ is the $\alpha$ component of phonon group velocity vector ${\boldsymbol{v}_{{{\lambda}}}}$, and $\tau_{{{\lambda}}, \alpha}$ is the phonon scattering time.  The phonon specific heat is obtained from the phonon vibrational frequencies as, $c_{{{\lambda}}} = \frac{\hbar\omega_{{{\lambda}}}}{V} \frac{\partial n^{o}_{{{\lambda}}}}{\partial T}$, where $n^o_{{{\lambda}}}$, $\hbar$, $\omega_{{{\lambda}}}$, $V$, and $T$ are the Bose-Einstein distribution, reduced Planck constant,   phonon frequency,  crystal volume,  and temperature. The phonon group velocities can be obtained as, ${\boldsymbol{v}_{{{\lambda}}}} = \frac{\partial \omega_{{\lambda}}}{\partial \boldsymbol{q}}$ and the phonon frequencies are obtained from the diagonalization of the dynamical matrix as, $\omega_{\lambda}^{2} {\boldsymbol{e}}_{\lambda} = {\boldsymbol{D}}_{{\boldsymbol{q}}} \cdot {\boldsymbol{e}}_{\lambda}$, where ${\boldsymbol{e}}_{\lambda}$ is the eigenvector and ${\boldsymbol{D}}_{{\boldsymbol{q}}}$ is the Dynamical matrix  whose elements, $D_{\boldsymbol{q}}^{3(b-1)+\alpha, 3(b^{'}-1) + \beta}$, are given by:
\begin{equation}
 \label{eqn_theory_dynamical}
 \begin{split}
 D_{\boldsymbol{q}}^{3(b-1)+\alpha, 3(b^{'}-1) + \beta} = \sum_{l^{'}} \frac{\Phi_{b0;b^{'}l^{'}}^{\alpha\beta} }{\sqrt{m_b m_{b^{'}}}}  
 e^{i[{\boldsymbol{q}}.( {\boldsymbol{r}}_{b^{'}l^{'}}  -  {\boldsymbol{r}}_{b0}  )] },
 \end{split}
\end{equation}
with summation running over all unit-cells in the lattice ($N$), and  $m_b$, $\boldsymbol{r}_{bl}$ being the mass and position vector  of atom $b$ in the in the $l^{th}$ unit-cell, and $\Phi_{ij}^{\alpha\beta}$ is the real-space ($ij, \alpha\beta$)-element of the  harmonic force constant matrix $\boldsymbol{\Phi}$.

%\subsection{Lowest-order Theory}
At the lowest level of theory, the $\boldsymbol{\Phi}$ used in Eqn.~\ref{eqn_theory_dynamical} are bare-harmonic force constant (as obtained from finite-difference or density functional perturbation theory and without renormalization). The phonon-phonon scattering rates at the lowest-order of theory are limited to three-phonon processes and are obtained as \cite{reissland1973, wallace1972, jain2020, godse2022}:
\begin{equation}
\begin{split}
 \label{eqn_rta_3ph}
 \frac{1}{\tau_{\lambda}^{3ph}} 
 =
  \sum_{\lambda_1, \lambda_2}
 \big\{
  {(n_{\lambda_1}  - n_{\lambda_2})}
 W^{+}
  +
  \frac{1}{2}
  (n_{\lambda_1} + n_{\lambda_2} +  1)
  W^{-}
   \big\},
 \end{split}
 \end{equation}
where $\boldsymbol{W}$ represents scattering probability matrix given by:
\begin{equation}
    \begin{split}
    \label{eqn_W_3ph}
W^{\pm}       
=
\frac{2\pi}{\hbar^2}
 \left|
 \Psi_{ {\lambda} (\pm{\lambda_{1}}) (-{\lambda_{2}}) }
 \right|^2
  \delta({\omega_{\lambda} \pm \omega_{\lambda_1} - \omega_{\lambda_{2}}}),
    \end{split}
\end{equation}
$\pm\lambda = (\pm\boldsymbol{q}, \nu)$,  and $\Psi_{ {\lambda} ({\lambda_{1}}) ({\lambda_{2}}) }$ are the Fourier transform of real-space cubic constants, $\Psi^{\alpha\beta\gamma}_{bl;b^{'} l^{'};b^{''} l^{''}}$, and are obtained as:
\begin{equation}
    \begin{split}
        \label{eqn_cubic_IFC}
\Psi_{ {\lambda} {\lambda_{1}} {\lambda_{2}} }
=
 \Psi_{ {\boldsymbol{q}} {\boldsymbol{q}^{'}} {\boldsymbol{q}^{''}} }^{\nu \nu^{'} \nu^{''}} = 
 N
 {\left(\frac{\hbar}{2N}\right)}^{\frac{3}{2}}
 \sum_{b} \sum_{b^{'} l^{'}}
\sum_{b^{''} l^{''}} 
\sum_{\alpha\beta\gamma} 
\Psi^{\alpha\beta\gamma}_{bl;b^{'} l^{'};b^{''} l^{''}}
\\
\times
\frac{
{{\tilde{e}}_{b\lambda}^{\alpha}}  
{{\tilde{e}}_{b^{'}{\lambda}^{'}}^{\beta}} 
{{\tilde{e}}_{b^{''}{\lambda}^{''} }^{\gamma}} }{\sqrt{ 
{m_b \omega_{\lambda}}  
{m_{b^{'}} \omega_{\lambda^{'}}}   
{m_{b^{''}} \omega_{\lambda^{''}}}   }}  
e^{[i( {{\boldsymbol{q}}^{'}}  \cdot{\boldsymbol{r}}_{0l^{'}} 
+    {{\boldsymbol{q}}^{''}}  \cdot{\boldsymbol{r}}_{0l^{''}} )]}.
    \end{split}
\end{equation}
The $\delta$ in Eqn.~\ref{eqn_cubic_IFC} represents the delta-function ensuring energy conservation and the summation is performed over phonon wavevectors satisfying crystal momentum conservation, i.e., $\boldsymbol{q} + \boldsymbol{q_1} + \boldsymbol{q_2} = \boldsymbol{G}$, where $\boldsymbol{G}$ is the reciprocal space lattice vector.

%\subsection{Higher-order Corrections: Four-phonon Scattering} 
At the higher-order of theory, the phonon-phonon scattering rates are obatined by considering both three-phonon and four-phonon processes. The scattering rates of phonon due to $2^{\text{nd}}$-order, i.e., four-phonon processes are obtained as \cite{reissland1973, feng2017, ravichandran2018, jain2020}:
\begin{equation}
 \begin{split}
  \label{eqn_rta_4ph}
  \frac{1}{\tau_{\lambda}^{4ph}}
 =
 \sum_{\lambda_{1}, \lambda_2, \lambda_3}
 \bigg\{
 \frac{1}{6}
 \Big\{
 \frac{ 
 n_{{\lambda_{1}}}  
 n_{{\lambda_{2}}} 
 n_{{\lambda_{3}}}  } 
 {
 n_{{\lambda_{}}}
 }
 W^{--}
 \Big\}
 + \\
\frac{1}{2}
 \Big\{
 \frac{ 
 (n_{{\lambda_{1}}}  + 1 )
 n_{{\lambda_{2}}} 
 n_{{\lambda_{3}}}  } 
 {
 n_{{\lambda_{}}}
 }
  W^{+-}
 \Big\}
  +
  \\
  \frac{1}{2}
 \Big\{
 \frac{ 
 (n_{{\lambda_{1}}}  + 1 )
 (n_{{\lambda_{2}}} +1)
 n_{{\lambda_{3}}}  } 
 {
 n_{{\lambda_{}}}
 }
  W^{++}
 \Big\}
 \bigg\},
\end{split}
\end{equation}
where $W^{\pm\pm}$ represents scattering probability matrix given by:
\begin{equation}
    \begin{split}
    \label{eqn_W_4ph}
W^{\pm\pm}       
=
\frac{2\pi}{\hbar^2}
 \left|
 \Xi_{ {\lambda} (\pm{\lambda_{1}}) (\pm{\lambda_{2}}) (-{\lambda_{2}}) }
 \right|^2
  \delta({
  \omega_{\lambda_{}} \pm 
  \omega_{\lambda_{1}} \pm 
  \omega_{\lambda_{2}} -
  \omega_{\lambda_{3}}
  }),
    \end{split}
\end{equation}
and $ \Xi_{ {\lambda} ({\lambda_{1}}) ({\lambda_{2}}) ({\lambda_{2}}) }$ are the Fourier transform of quartic force constants, $\Xi_{ijkl}^{\alpha\beta\gamma\delta}$, and are obtained as:
\begin{equation}
    \begin{split}
        \label{eqn_quartic_IFC}
\Xi_{ {\lambda} {\lambda_{1}} {\lambda_{2}} {\lambda_{2}} }
=
 \Xi_{ {\boldsymbol{q}} {\boldsymbol{q}^{'}} {\boldsymbol{q}^{''}}  {\boldsymbol{q}^{'''}} }^{\nu \nu^{'} \nu^{''} \nu^{'''} } = 
 N
 {\left(\frac{\hbar}{2N}\right)}^{{2}}
 \sum_{bb^{'}b^{''}b^{'''}   } \sum_{l^{'}l^{''}l^{'''}}
\sum_{\alpha\beta\gamma\delta} \\
\Xi^{\alpha\beta\gamma\delta}_{bl;b^{'} l^{'};b^{''} l^{''};b^{'''} l^{'''}}
\frac{
{{\tilde{e}}_{b\lambda}^{\alpha}}  
{{\tilde{e}}_{b^{'} \lambda^{'}}^{\beta}} 
{{\tilde{e}}_{b^{''}\lambda^{''}}^{\gamma}} 
{{\tilde{e}}_{b^{'''} \lambda^{'''}}^{\delta}} }
{\sqrt{ 
{m_b \omega_{\lambda}}  
{m_{b^{'}} \omega_{\lambda^{'}}}   
{m_{b^{''}} \omega_{\lambda^{''}}}   
{m_{b^{'''}} \omega_{\lambda^{'''}}} 
}}  
\\
e^{[i( {{\boldsymbol{q}}^{'}}  \cdot{\boldsymbol{r}}_{0l^{'}} 
+    {{\boldsymbol{q}}^{''}}  \cdot{\boldsymbol{r}}_{0l^{''}} 
+    {{\boldsymbol{q}}^{'''}}  \cdot{\boldsymbol{r}}_{0l^{'''}}   )]}.
    \end{split}
\end{equation}
Similar to Eqns.~\ref{eqn_W_3ph} and \ref{eqn_cubic_IFC}, the summations in Eqn.~\ref{eqn_quartic_IFC} are performed over phonon wavevectors satisfying crystal momentum conservation and $\delta$ ensures energy conservation during four-phonon scattering event. 

%\subsection{Higher-order Corrections: Phonon renormalization}
Further, at the higher-order of theory, the phonon frequencies and eigenvectors are obtained from Eqn.~\ref{eqn_theory_dynamical} using the renormalized harmonic force constants, $\Phi^{c, \alpha\beta}_{ij}$, which are obtained as \cite{wallace1972, ravichandran2018}: 
\begin{eqnarray}
\begin{split}
\Phi^{c, \alpha\beta}_{ij} = 
    \Phi^{o, \alpha\beta}_{ij} + 
    \frac{\hbar}{4N}\sum_{l^{'''}l^{''''}} \sum_{b^{'''}b^{''''}}\sum_{\gamma\delta}\sum_{\boldsymbol{q}\nu}
    \Xi_{ijkl}^{\alpha\beta\gamma\delta}
    \\
    \frac{{{\tilde{e}^\gamma}_{b^{'''}\lambda}}
    {{\tilde{e}^{\dagger\delta}}_{b^{''''} \lambda}}}
    {\omega_{\lambda}\sqrt{m_{b^{'''}} m_{b^{''''}}}}
    (2n_{\lambda}+1)
    e^{i\boldsymbol{q}\cdot{ (\boldsymbol{r}_{0l^{'''}} - \boldsymbol{r}_{0l^{''''}}) }}, 
    \label{eqn_renorm}
\end{split}
\end{eqnarray}
where $\Phi^{o, \alpha\beta}_{ij}$ represent the raw harmonic force constants (as obtained from ab-initio calculations).  

%\subsection{Multichannel Thermal Transport}
Finally, to account for multi-channel thermal transport, the contribution of coherent channel, $k_{\alpha}^c$, can be obtained using the off-diagonal terms of the velocity operator as \cite{simoncelli2019}:
\begin{equation}
\begin{split}
    \label{eqn_coherent}
k^c_{\alpha} 
 = 
 \frac{\hbar^2}{k_BT^2}
 \frac{1}{VN}
 \sum_{\boldsymbol{q}}
 \sum_{(\nu_{} \ne \nu_{1})}
 \frac{ \omega_{\boldsymbol{q}\nu_{}} +  \omega_{\boldsymbol{q}\nu_{1}} }{2}
 V^{\alpha}_{\boldsymbol{q}, \nu\nu_{1}}
 V^{\alpha}_{\boldsymbol{q}, \nu_1\nu_{}} 
 \times
 \\
 \frac{
  \omega_{\boldsymbol{q}\nu_{}}  n_{\boldsymbol{q}\nu_{}} (n_{\boldsymbol{q}\nu_{}} + 1)
  +
  \omega_{\boldsymbol{q}\nu_{1}}  n_{\boldsymbol{q}\nu_{1}} (n_{\boldsymbol{q}\nu_{1}} + 1)
 }
 {
 4(\omega_{\boldsymbol{q}\nu_{}} - \omega_{\boldsymbol{q}\nu_1})^2 
 + (\Gamma_{\boldsymbol{q}\nu_{}} + \Gamma_{\boldsymbol{q}\nu_{1}} )^2
 }
 (\Gamma_{\boldsymbol{q}\nu_{}} + \Gamma_{\boldsymbol{q}\nu_{1}}),
\end{split}
\end{equation}
where $\Gamma_{\boldsymbol{q}\nu_{}}$ is the phonon linewidth ($\Gamma_{\boldsymbol{q}\nu_{}} = 1/\tau_{\boldsymbol{q}\nu_{}}$) and $V^{\alpha}_{\boldsymbol{q}, \nu\nu_{1}}$ is the $\alpha$-component of velocity operator and can be obtained as:
\begin{equation}
    {\boldsymbol{V}}_{\boldsymbol{q}, \nu\nu_{1}} = 
    \frac{1}{2\sqrt{ \omega_{\boldsymbol{q}\nu_{}} \omega_{\boldsymbol{q}\nu_{1}} }} 
    \bra{ {{{\boldsymbol{e}}}_{\boldsymbol{q}\nu}}} 
    \frac{\partial \boldsymbol{D}_{\boldsymbol{q}} }{\partial \boldsymbol{q}} 
    \ket{ {{{\boldsymbol{e}}}_{\boldsymbol{q}\nu_{1}}}},
\end{equation}
using the Wallace/smooth representation of the dynamical matrix (Eqn.~\ref{eqn_theory_dynamical}) \cite{simoncelli2021}. Further details on these methods are available in Refs.~\cite{jain2020,  godse2022}. 
The computational details and the convergence of predicted results with the choice of numerical parameters are reported in Secs.~S1-S3 in the Supplementary Information (SI).

\label{sec_higherTheory}
\begin{figure}
\begin{center}
\epsfbox{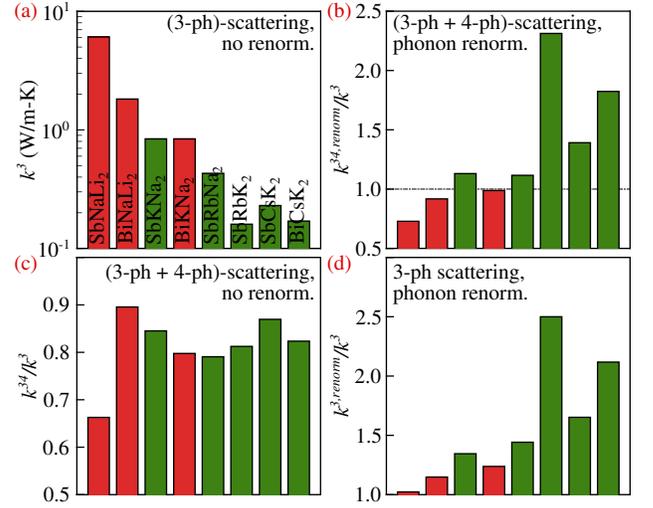}
\end{center}
\caption{The thermal conductivity of considered compounds at 300 K as obtained by considering (a) lowest-order theory, $k^{3}$, and (b) higher-order four-phonon scattering and quartic renormalization corrections, $k^{34,renorm}$.  The relative importance of (c) four-phonon scattering, $k^{34}/k^{3}$, and (d) quartic renormalization, $k^{3,renorm}/k^{3}$, on the predicted thermal conductivity. The materials for which thermal conductivity increases (decreases) with higher-order theory are colored green (red). }
\label{fig_conductivity}
\end{figure}

By considering only the particle-like transport channel, at a temperature of 300 K, the thermal conductivities of considered compounds vary between $0.16$-$6.08$ $\text{W/m-K}$  using the lowest-order theory as reported in Fig.~\ref{fig_conductivity}(a). On inclusion of higher-order corrections in Fig.~\ref{fig_conductivity}(b), the thermal conductivity decreases for three compounds and increases for five compounds. The maximum decrease (increase) is for $\text{SbNaLi}_2$ ($\text{SbRbK}_2$), for which thermal decreases by 27\% (increases by more than a factor of two) with the inclusion of higher-order corrections.   To understand the origin of this puzzling role of higher-order corrections on the predicted thermal conductivity, the effect of  four-phonon scattering and phonon renormalization are considered separately.

\begin{figure}
\begin{center}
\epsfbox{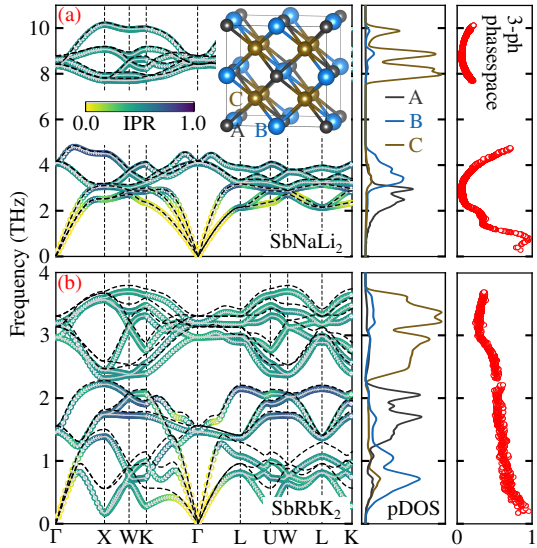}
\end{center}
\caption{The representative phonon dispersion of materials with (a) frequency bandgaps   and (b) weakly bonded atoms.  The grey solid and black dashed lines represent non-renormalized and renormalized (at 300 K) results. The markers are colored according to the inverse participation ratio of modes which varies between 0 and 1 for localized and delocalized modes.}
\label{fig_harmonic}
\end{figure}

\begin{figure}
\begin{center}
\epsfbox{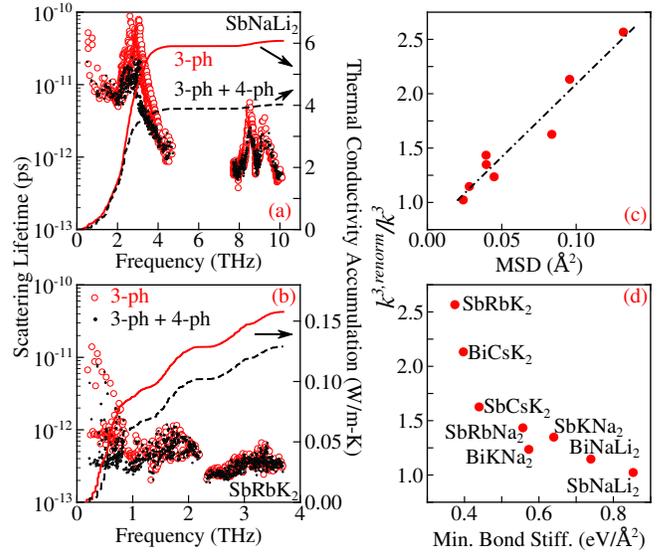}
\end{center}
\caption{The phonon scattering lifetimes obtained by considering only three-phonon scattering and by considering both three- and four-phonon scattering for (a) $\text{SbNaLi}_2$ and (b) $\text{SbRbK}_2$. The thermal conductivity enhancement of considered compounds with phonon renormalization plotted against (c) thermal MSD and (d) effective bond stiffness. The reported MSD (bond stiffness) in (c) [(d)] are the maximum (minimum) of all atoms in a considered compound.}
\label{fig_analysis}
\end{figure}

%\subsubsection{Four-phonon Scattering}
Compared to the lowest-order theory, the inclusion of four-phonon scattering processes decreases the thermal conductivity for all considered compounds due to an additional phonon scattering channel. The extent to which the phonon thermal conductivity is affected depends on two factors, (\roml{1}) how much was three-phonon scattering to start with and  (\roml{2}) how strong is four-phonon scattering. The extent of three-phonon scattering, in turn, is dependent on the number of three-phonon processes which are able to satisfy the scattering selection rules and the strength of each of these scattering processes. The number of three-phonon processes which can satisfy the crystal momentum and energy conservation selection rules can be obtained from phonon spectrum and are characterizable using the three-phonon scattering phase space (see Fig.~\ref{fig_harmonic}). For instance, amongst the two representative phonon spectrums presented in Fig.~\ref{fig_harmonic}, the presence of frequency gaps in spectrum of $\text{SbNaLi}_2$ makes it difficult  to satisfy the energy conservation selection rule (see details in Sec.~S3) and results in the reduction of three-phonon scattering phase space (to almost zero) for several mid- and high-frequency optical phonons. This non-scattering of optical phonons via three-phonon processes gets reflected in their high contribution to thermal transport at the lowest-order of theory. With inclusion of higher-order four-phonon scattering, these otherwise unscattered phonons are able to undergo scattering which results in their reduced contribution to thermal transport \cite{ph4PS}. In the particular case of $\text{SbNaLi}_2$ at 300 K, the contribution of mid-frequency optical phonons (2-4 THz) is reduced from $4.3$ $\text{W/m-K}$ to $2.4$ $\text{W/m-K}$ with inclusion of four-phonon scattering [Fig.~\ref{fig_analysis}(a)] which corresponds to a 34\% reduction in the total predicted thermal conductivity.

The strength of individual three-phonon and four-phonon scattering processes depends on the anharmonicity experienced by atoms. For materials with strongly bonded atoms, the atoms move very little around their equilibrium positions and are limited to the harmonic part of the potential energy surface. In contrast, for materials with weakly-bonded atoms, the atoms are able to move far away from their equilibrium positions and experience the more anharmonic part of the potential energy surface. In such materials,  the effect of four-phonon scattering is strongest on modes dominated by loosely bound atoms (which happens to be the heat-carrying acoustic modes). For instance, in $\text{SbRbK}_2$, the  acoustic modes are dominated by weakly bonded Rb atoms and have frequencies up to $1.5$ $\text{THz}$. In contrast, for $\text{SbNaLi}_2$, the corresponding modes have contribution from strongly-bonded Sb-atoms and have frequencies up to 2.5 THz. As a result, while the contribution of sub-THz modes reduces by 29\% with inclusion of four-phonon scattering in $\text{SbRbK}_2$, the corresponding reduction is less than 8\% for modes with frequencies up to 2 THz in $\text{SbNaLi}_2$ [Fig.~\ref{fig_analysis}(b)].

%\subsubsection{Phonon Renormalization} 

With phonon renormalization, the heat-carrying acoustic phonons become stiff in all considered materials resulting in an increase in the predicted thermal conductivity [Fig.~\ref{fig_conductivity}(d)]. The extent of renormalization is proportional to  (a) the thermal MSD of atoms  and (b) the strength of quartic force constant (Eqn.~\ref{eqn_renorm}). The strength of quartic force constants or anharmonicity is, in general,  dependent on atomic displacements around their equilibrium positions. As such, the enhancement in thermal conductivity  with renormalization is expected to have a strong dependence on thermal MSD. This is reported in Fig.~\ref{fig_analysis}(d) where a linear dependence is obtained between the thermal conductivity enhancement and MSDs. Since large MSDs are  a manifestation of weak atomic bonding, the thermal conductivity enhancement also correlate strongly with the effective bond stiffness [Fig.~\ref{fig_analysis}(d)] \cite{bond_stiffness}.

%\subsubsection{Descriptors for Higher-order Corrections}
\begin{figure}
\begin{center}
\epsfbox{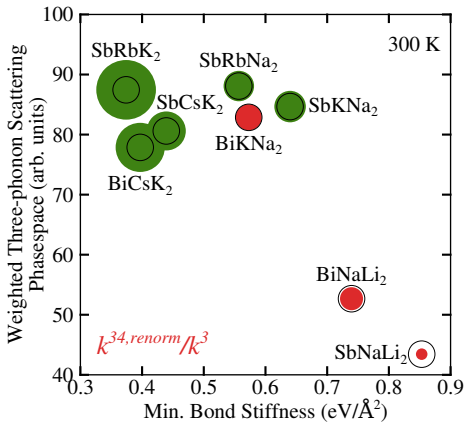}
\end{center}
\caption{The change in thermal conductivity of considered compounds with higher-order corrections as summarized using bond stiffness and scattering phase space. The materials for which thermal conductivity decreases (increases) with higher-order corrections are highlighted using red (green) circles. The size of circles is proportional to the fractional change in thermal conductivity ad black circles represent  no change in thermal conductivity. }
\label{fig_descriptor}
\end{figure}

Now that it is established that the three-phonon scattering phase space and effective bond stiffness are instrumental in describing the individual effects of four-phonon scattering and phonon renormalization,  the otherwise puzzling thermal conductivity data of Fig.~\ref{fig_conductivity}(b) is re-plotted in Fig.~\ref{fig_descriptor} as a function of two simple material descriptors, namely, weighted three-phonon scattering phase space and effective bond stiffness. The weighted three-phonon scattering phase space, $\overline{\gamma}$, is obtained as $\overline{\gamma} = \frac{ \sum_{{\boldsymbol{q}}} \sum_{\nu} c_{{\boldsymbol{q}}\nu} v_{{\boldsymbol{q}}\nu, \alpha}^{2} \gamma_{{\boldsymbol{q}}\nu} }{ \sum_{{\boldsymbol{q}}} \sum_{\nu} c_{{\boldsymbol{q}}\nu} v_{{\boldsymbol{q}}\nu, \alpha}^{2} }$, where $\gamma_{{\boldsymbol{q}}\nu} $ represents scattering phase space of mode ${{\boldsymbol{q}}\nu} $.  As can be seen from Fig.~\ref{fig_descriptor}, these descriptors are able to characterize the role of higher-order theory on predicted thermal conductivity and suggest that:  (\roml{1}) for materials with weak inter-atomic bonding, the thermal conductivity increases with higher-order corrections, (\roml{2}) for materials  with reduced three-phonon scattering phase space, the thermal conductivity decreases with higher-order corrections, and (\roml{3}) for materials with stiff bonds and high three-phonon scattering phase-space, there is no significant change in thermal conductivity.  As such, these simple (based on bare harmonic properties) and yet powerful descriptions can be used for accelerated discovery/screening of ultra-low thermal conductivity materials without requiring computationally expensive calculations.

%\subsection{Multi-channel Thermal Transport}
\begin{figure}
\begin{center}
\epsfbox{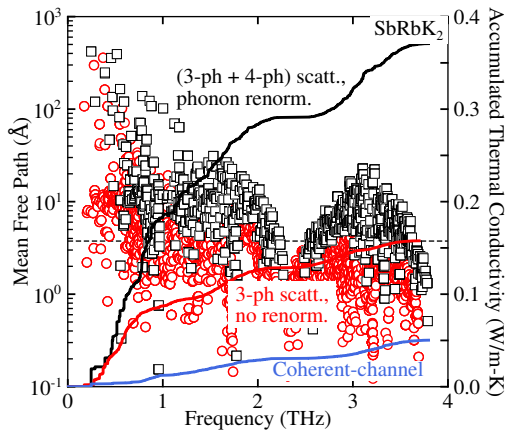}
\end{center}
\caption{The mode-dependent contribution  to thermal conductivity of $\text{SbRbK}_2$ from particle-like phonon (red and black lines) and wave-like coherent (purple line) transport channels. The red and black datasets correspond to results obtained using the lowest-order theory and higher-order theory. The dashed horizontal line represents the Ioffe-Regel limit \cite{regel1960, ioffe_regel_limit}.}
\label{fig_coherent}
\end{figure}

Using the lowest-order transport theory, the predicted thermal conductivity of two compounds is lower than $0.2$ $\text{W/m-K}$ at a temperature of 300 K which is comparable to the reported lowest thermal conductivity  of solids in literature \cite{mukhopadhyay2018, jain2020, godse2022}. For such low thermal conductivity materials, the phonon mean free paths could become shorter than the Ioffe-Regel limit, and in such cases, an additional contribution from wave-like coherent transport channel becomes significant towards the thermal transport \cite{regel1960, mukhopadhyay2018, simoncelli2019}. To check for this, the phonon mean free paths are compared against the Ioffe-Regel limit \cite{ioffe_regel_limit} for the lowest thermal conductivity (as obtained from the lowest-order theory) compound in Fig.~\ref{fig_coherent}.

As can be seen from Fig.~\ref{fig_coherent}, while  the phonon mean free paths are shorter than the Ioffe-Regel limit  for the majority of the  phonons using the lowest-order theory, the mean free paths increase to values larger than the Ioffe-Regel limit on the inclusion of higher-order corrections. As such, the particle-like phonon picture is valid for the majority of the phonons and the contribution of wave-like coherent transport channel is  minimal (less than 13\% for $\text{SbRbK}_2$ using the higher-order theory as opposed to more than 30\% using the lowest-order theory). This validity of particle-like phonon picture on the incorporation of higher-order corrections and its failure at lowest-order of theory highlights the importance of describing correct phonon-properties  in multi-channel transport models to correctly capture the underlying thermal transport physics.

%\section{Conclusions}
In summary, by investigating the thermal transport in eight ternary semiconducting solids using the higher-order transport theory, the weighted three-phonon scattering phase space and effective bond stiffness are identified as simple material descriptors to characterize the puzzling role of higher-order four-phonon scattering and phonon renormalization on the predicted thermal conductivity of materials. The higher-order corrections are essential for the correct description of thermal transport physics in ultra-low thermal conductivity solids and the use of the lowest-order theory erroneously suggests multi-channel thermal transport in such materials.

\section{Acknowledgement}
The author acknowledges the financial support from IRCC-IIT Bombay and the National Supercomputing Mission, Government of India (Grant Number: DST/NSM/R\&D-HPC-Applications/2021/10).  The calculations are carried out on SpaceTime-II supercomputing facility of IIT Bombay and PARAM Sanganak supercomputing facility of IIT Kanpur.

\bibliography{references}

\begin{thebibliography}{27}
\bibitem [{ph4()}]{ph4PS}%
  \BibitemOpen
  \href@noop {} {}\bibinfo {note} {Note that as presented in Sec.~S5, the
  number of four-phonon processes, i.e., four-phonon scattering phase space is
  non-zero for all modes and is minimally affected by frequency gaps in the
  phonon spectrum.}\BibitemShut {Stop}%
\bibitem [{bon()}]{bond_stiffness}%
  \BibitemOpen
  \href@noop {} {}\bibinfo {note} {The effective stiffness is obtained by
  displacing atoms by a small amount (up to $0.1$ $\text{\AA}$) around their
  equilibrium positions to sample the effective potential energy
  surface.}\BibitemShut {Stop}%
\end{thebibliography}
\end{document}